\documentclass[12pt]{article}
\usepackage{amssymb}

\begin{document}
\begin{titlepage}




{\hbox to\hsize{\hfill 
March 2016 }}

\bigskip \vspace{3\baselineskip}

\begin{center}

{\bf \Large 

Natural Inflation with Hidden Scale Invariance }

\vskip 0.4cm

\bigskip

\bigskip

\bigskip

\bigskip

{\bf Neil D. Barrie, Archil Kobakhidze and Shelley Liang \\}

\smallskip

\bigskip

{ \small \it

ARC Center of Excellence for Particle Physics at the Terascale, \\
School of Physics, The University of Sydney, NSW 2006, Australia \\
E-mails: neil.barrie, archil.kobakhidze, shelley.liang@sydney.edu.au
\\}

\bigskip

\bigskip

\bigskip

\bigskip

\bigskip

{\large \bf Abstract}

\end{center}

\noindent
We propose a new class of natural inflation models based on a hidden scale invariance. In a very generic Wilsonian effective field theory with an arbitrary number of scalar fields, which exhibits scale invariance via the dilaton, the potential necessarily contains a flat direction in the classical limit. This flat direction is lifted by small quantum corrections and inflation is realised without need for an unnatural fine-tuning. 
In the conformal limit, the effective potential becomes linear in the inflaton field, yielding to specific predictions for the spectral index and the tensor-to-scalar ratio, being respectively: $n_s-1\approx -0.025\left(\frac{N_{\star}}{60}\right)^{-1}$ and $r\approx 0.0667\left(\frac{N_{\star}}{60}\right)^{-1}$, where $N_{\star}\approx 30-65$ is a number of efolds during observable inflation. This predictions are in reasonable agreement with cosmological measurements. Further improvement of the accuracy of these measurements may turn out to be critical in falsifying our scenario.      
\end{titlepage}

\paragraph{Introduction.}
Cosmic inflation is an attractive paradigm for the very early universe that resolves some outstanding puzzles of the standard hot Big Bang cosmology, such as the horizon and flatness problems \cite{Guth:1980zm, Linde:1981mu, Albrecht:1982wi} (for important precursor works see also \cite{Kazanas:1980tx, Starobinsky:1980te, Sato:1981ds}). In addition, it provides a natural mechanism for generation of nearly scale-invariant inhomogeneities through the quatum fluctuations of the inflaton field, that at later stages result in the observed large scale structure of the universe \cite{Mukhanov:1981xt}. Observations on Cosmic Microwave Background (CMB) radiation and the large scale structure provide a strong support for cosmic inflation. 

The basic theory of inflation involves a scalar field, the inflaton ($\varphi$), which slowly rolls down the potential hill. In order to reproduce the CMB anisotropy measurements \cite{Ade:2015lrj} and satisfy the requirement of sufficient inflation, the scale that defines the height of the inflaton potential must be many orders of magnitude smaller than the scale that defines its width, that is, the potential must be very flat.  To maintain the hierarchy between these two different scales under the quantum corrections a precise adjustment of couplings is typically required. This is known as the fine-tuning problem of inflation. 

Another potential source that may destabilise the delicate balance between the height and slope of the inflaton potential 
is higher order operators, which start to contribute significantly for large variations of the inflaton field during inflation. The effective field theory approximation, which favours $|\varphi|\ll M_P$, breaks down in such cases and inflationary predictions become unreliable.

A class of natural inflation models has been suggested in \cite{Freese:1990rb} as a symmetry-motivated solution to the above fine-tuning problem. The inflaton in this class of models is a pseudo-Goldstone boson of some spontaneously broken anomalous global symmetry. The flatness of the pseudo-Goldstone potential is guaranteed by an approximate shift symmetry, although the underlying global symmetry may be the subject of large explicit breaking by non-renormalisable operators supposedly induced via quantum gravity. It seems, however, that the simplest models of natural inflation are now disfavoured at 95\% CL \cite{Ade:2015lrj}.       

In some earlier works \cite{Wetterich:1983bi, Bardeen:1995kv} and more recently in \cite{Foot:2013hna, Kobakhidze:2014afa} the scale invariance was advocated as a possible symmetry which is also capable of explaining the hierarchy of different scales without fine-tuning. A variety of specific scale-invariant inflationary models have been presented in recent years \cite{GarciaBellido:2011de, Kallosh:2013hoa, Salvio:2014soa, Ellis:2014gxa, Kannike:2014mia, Csaki:2014bua, Kannike:2015apa,Ozkan:2015kma,Kannike:2015fom, Farzinnia:2015fka,Rinaldi:2015uvu}. In \cite{Kallosh:2013hoa} a universality class of models has been identified within the conformal supergravity framework [see also Ref. \cite{Ellis:2014gxa}]. The importance of the underlying scale invariance for natural inflation models has also been stressed in \cite{Csaki:2014bua}. 

In this paper we would like to propose a new class of natural inflation models based on a hidden scale invariance, realised through the pseudo-Goldstone boson of a spontaneously broken anomalous scaling symmetry, the dilaton.  Our key observation is that in a very generic scale-invariant model, with an arbitrary number of scalar fields and non-renormalizable operators in the scalar potential, there always exists a direction in field space which is absolutely flat in the classical limit. This flat direction is lifted upon quantum corrections being taken into account. Inflation proceeds along this direction, while other fields reside in their respective (meta)stable minima.  As will be shown below, 
in the conformal coupling limit within the leading perturbative approximation, the generic model is reduced to a one-field model with a potential linear in the inflaton field, $V(\varphi)\sim \varphi$, with the linear term being radiatively induced. In this regime, the model predicts a characteristic relation between the spectral index $n_s$ and the tensor-to-scalar ratio $r$:
\begin{eqnarray}
n_s\approx 1-\frac{3}{8}r
\label{1} 
\end{eqnarray}
This relation is in a reasonable agreement with the currently available data \cite{Ade:2015lrj, Huang:2015cke}. Further improvement on the accuracy of $n_s$ and/or $r$ measurements may confirm or falsify our scenario.   

\paragraph{Description of the model.}
Consider a Wilsonian effective  field theory that describes the Standard Model, or its extension, coupled to gravity at an ultraviolet scale $\Lambda$: 
\begin{equation}
S_{\Lambda}=\int dx^4 \sqrt{-g}\left[\left(\frac{M_P^2}{2}+\sum_{i=1}^N\xi_i(\Lambda)\phi_i^2\right)R-\frac{1}{2}\sum_{i=1}^N \partial_{\mu}\phi_i\partial^{\mu}\phi_i -V(\phi_i)+...\right]~,
\label{3}
\end{equation}
where $M_P\approx 2.4\cdot 10^{18}$ GeV and we use the mostly positive signature for the metric tensor. Here we have displayed only the scalar sector, which comprises of a set of $N$ scalar fields $\{\phi_i\}$ $(i=1,2,...,N)$ that includes the Standard Model Higgs boson. The scalar potential $V(\phi_i)$ is a generic polynomial of the scalar fields $\{\phi_i\}$ respecting the relevant symmetries of the theory:
\begin{equation}
V(\phi_i)=\sum_{\{i_n\}}\lambda_{i_1,...,i_n}(\Lambda)\phi_{i_1}...\phi_{i_n}~. 
\label{4}
\end{equation}   
 where $\lambda_{i_1,...,i_n}(\Lambda)$ is a coupling of mass dimension $(4-n)$ defined at the Wilsonian cut-off $\Lambda$, while $\xi_i(\Lambda)$ is a dimensionless non-minimal coupling of the scalar field $\phi_i$ to gravity. The scale invariance is explicitly broken in (\ref{3}) by the ultraviolet cut-off $\Lambda$, the Einstein-Hilbert term $\sim M_P^2 R$ and dimensionful couplings $\sigma_{i_1,...,i_n}$ ($n\neq 4$).    

We suppose that the underlying theory exhibits a hidden (spontaneously broken) scale invariance, which in the effective low-energy theory   is implemented in the (nonlinear) pseudo-Goldstone boson, the dilaton $\chi$. A simple way to incorporate the dilaton field $\chi$ is to rescale the dimensionful parameters in (\ref{3}) by the respective powers of $\chi /f$, $f$ being the dilaton ``decay constant". More specifically: 
\begin{eqnarray}
\Lambda\to \Lambda \frac{\chi}{f}\equiv \lambda \chi~,~~M_P^2 \to M_P^2 \left(\frac{\chi}{f}\right)^2\equiv \xi \chi^2~,~~ \\ 
\lambda_{i_1,...,i_n}(\Lambda) \to \lambda_{i_1,...,i_n}(\Lambda \chi /f) \left(\frac{\chi}{f}\right)^{4-n}\equiv \sigma_{i_1,...,i_n}(\lambda \chi)\chi^{4-n}
\label{5}
\end{eqnarray}
Thus, instead of  (\ref{3}) we consider a new action:
 \begin{eqnarray}
S_{\lambda\chi}&=&\int dx^4 \sqrt{-g}\left[\left(\xi \chi^2+\sum_{i=1}^N \xi_i(\lambda\chi)\phi_i^2\right)R-\frac{1}{2}\partial_{\mu}\chi\partial^{\mu}\chi-\frac{1}{2}\sum_{i=1}^N \partial_{\mu}\phi_i\partial^{\mu}\phi_i -V(\phi_i, \chi)+...\right]~, \nonumber \\
V(\phi_i, \chi)&=&\sum_{\{i_n\}}\sigma_{i_1,...,i_n}(\lambda \chi)~\chi^{(4-n)}\phi_{i_1}...\phi_{i_n}~.
\label{6}
\end{eqnarray}
 This action is manifestly scale invariant in the classical limit, the scale invariance being broken at the quantum level through the renormalisation group (RG) running of the couplings, i.e., $\frac{\partial \sigma_{i_1,...,i_n}}{\partial \chi}\neq 0$, etc. 
  
It is convenient to use a `hyperspherical' representation for the set of scalar fields $\{\phi_i, \chi\}$:
\begin{eqnarray}
\phi_i &=&\rho\cos\left(\theta_{i}\right)\prod_{k=1}^{i-1}\sin\left(\theta_k\right)~,~~(i=1,2,...,N) \nonumber \\
\chi &=&\rho \prod_{k=1}^{N}\sin\left(\theta_k\right)~.
\label{7}
\end{eqnarray}
Expressing the action (\ref{6}) through the fields in the above representation, we observe that the modulus field $\rho$ factors out. That is, the first term in the action and the scalar potential presented in Eq. (\ref{6}) can be written as $\sim \rho^2 \zeta (\theta_i)R$ and $\sim \rho^4 U(\theta_i)$, respectively, in which
\begin{eqnarray}
\label{8}
\zeta(\theta_{i})&=&\xi(\lambda\chi) \prod_{k=1}^{N}\sin^2\left(\theta_k\right)+\sum_{i=1}^{N}\xi_i(\lambda\chi)\cos^2\left(\theta_{i}\right)\prod_{k=1}^{i-1}\sin^2\left(\theta_k\right)~, \\
U(\theta_{i})&=&\prod_{k=1}^{N}\sin^{4-n}\left(\theta_k\right)\sum_{\{i_n\}}\sigma_{i_1,...,i_n}(\lambda\chi)\cos\left(\theta_{i_1}\right)
\prod_{k=1}^{i_1-1}\sin\left(\theta_k\right)...\cos\left(\theta_{i_n}\right)
\prod_{k=1}^{i_n-1}\sin\left(\theta_k\right)~.
\label{9}
\end{eqnarray}    

We further assume that $\theta_i$ fields are relaxed in their stable or sufficiently long-lived (with lifetime longer than the duration of the observable inflation) minima $\langle\theta_i \rangle = \theta_i^c$  at very early stages in the evolution of the universe. Hence,  their dynamics is of no interest to us in what follows and, instead of the full action (\ref{6}), we consider the following reduced one:
\begin{eqnarray}
\label{10}
\bar S_{\rho}&=&\int dx^4 \sqrt{-g}\left[\zeta(\rho) \rho^2 R-\frac{1}{2}\partial_{\mu}\rho\partial^{\mu}\rho-V(\rho) \right]~, \\
V(\rho)&=&\sigma (\rho)\rho^4~, 
\label{11}
\end{eqnarray}
where $\zeta\equiv \zeta(\theta_i^c)$ and $\sigma\equiv U(\theta_i^c)$. Hence, we arrive at an effective single-field model with a quartic potential and non-minimal coupling \cite{Okada:2010jf}, but without the standard Einstein-Hilbert term. It resembles also the large field limit of the Higgs inflation model \cite{Bezrukov:2007ep}. 

In order to reproduce the Einstein-Hilbert term in (\ref{10}) the modulus field $\rho$ has to develop non-zero vacuum expectation value, $\langle\rho \rangle\equiv \rho_0$. If the vacuum configuration $\{\rho_0, \theta_i^c\}$ describes the current vacuum state of the universe, than  $\rho_0= \frac{M_P}{\sqrt{2\zeta(\rho_0)}}$ with $\zeta (\rho_0)\equiv \zeta_0 >0$. Furthermore,  the vacuum energy density, $\frac{\sigma(\rho_0)M_P^4}{4\zeta_0^2}$, in this case must be vanishingly small to satisfy the observations. That is, the scalar potential must be tuned so that $\sigma(\rho_0)\equiv \sigma_0 \sim 12\zeta_0^2H_0^2/M_P^2\approx 0$, where $H_0$ is the present value of the Hubble parameter. However, inflation may end in a metastable state, which subsequently decays into the current vacuum state. Hence, we keep $\rho_0$ and $\sigma_0$ as a free parameters.    

The $\rho-$dependence of dimensionless couplings in Eq. (\ref{10}, \ref{11}) are determined by computing the quantum-corrected effective potential. We use the  closed form effective potential computed in Ref. \cite{Odintsov:1993rt} to obtain in the 1-loop approximation:
\begin{eqnarray}
\label{12}
\zeta (\rho)&=&\zeta_0+\frac{(12\zeta_0+1)\sigma_0}{8\pi^2}\ln\left(\frac{\rho}{\rho_0}\right) ~,\\
\label{13}
\sigma (\rho)&=&\sigma_0+\frac{9\sigma_0^2}{2\pi^2}\ln\left(\frac{\rho}{\rho_0}\right) ~.
\end{eqnarray} 
In the classical limit $\zeta$ and $\sigma$ are $\rho$-independent constants and the action (\ref{10}) is scale-invariant. The field $\rho$ represents a flat direction, i.e., the potential (\ref{11}) is constant for any value of $\rho$ in the Einstein frame. Furthermore, for the special value $\zeta_0 = -1/12$ (the conformal coupling), $\rho$ is a fictitious degree of freedom which disappears from action in the Einstein frame.  In this case, the action (\ref{10}) is in fact describes pure Einstein gravity with a cosmological constant.

The classical scale invariance is broken by radiative corrections, which is illustrated in the $\rho$ dependence of couplings, Eqs. (\ref{12},\ref{13}). Note that $\sigma_0 \to 0$ is a conformal fixed-point of the theory, since the $\rho$ dependence disappears in Eqs. (\ref{12},\ref{13}) in this limit. The conformal coupling $\zeta_0 = -1/12$ is also a fixed-point as $\zeta(\rho)=\zeta_0$. Hence,  having $\sigma$ small or $\zeta$ close to $-1/12$ near the respective fixed points  is natural in the technical sense. All these attractive features motivate us to consider scale invariance as an essential symmetry for natural inflation, with $\rho$ being the inflaton field.

\paragraph{Predictions of the model.}
To compute inflationary observables we first take the action (\ref{10}) to the Einstein frame via Weyl rescaling: 
\begin{equation}
g_{\mu\nu} \to \Omega^2 g_{\mu\nu}~,~~\Omega^2= \frac{2\zeta \rho^2}{M_P^2}
\label{14}
\end{equation}
We also bring the kinetic term for the inflaton field $\rho$ to the canonical form by making the following field redefinition:
\begin{equation}
\rho = \rho_0\exp \left( \frac{\sqrt{\tilde \zeta}}{M_P}\varphi\right)~,
\label{15}
\end{equation}
where $\tilde \zeta = \frac{2\zeta}{1+12\zeta}$ with $\zeta>0$ or $\zeta<-1/12$.  With this the action (\ref{10}) in the Einstein frame reads: 
\begin{eqnarray}
\label{16}
\bar S_{\varphi}&=&
\int dx^4 \sqrt{-g}\left[\frac{M_P^2}{2} R-\frac{1}{2}\partial_{\mu}\varphi\partial^{\mu}\varphi-V(\varphi) \right]~, \\
V(\rho(\varphi))&=&\frac{M_P^4}{4}\frac{\sigma (\rho(\varphi))}{\zeta^2(\rho(\varphi))}~. 
\label{17}
\end{eqnarray}

Given the potential (\ref{17}), the slow roll parameters can be computed using:
\begin{eqnarray}
\label{18}
\epsilon_{\star}&\equiv&\frac{M_P^2}{2}\left.\left(\frac{V_{\varphi}}{V}\right)^2\right \vert_{\varphi=\varphi_{\star}}~, \\
\eta_{\star}& \equiv &\left. M_P^2\frac{V_{\varphi\varphi}}{V}\right \vert_{\varphi=\varphi_{\star}}
~.
\label{19}
\end{eqnarray}
The power spectrum of scalar perturbations, $P_{s}$, the tensor-to-scalar ratio, $r$, and the spectral index $n_s$ are then given by:
 \begin{eqnarray}
\label{20}
P_s&=&\frac{1}{24\pi^2 M_P^4}\frac{V_{\star}}{\epsilon_{\star}}~, \\
\label{21}
r&=&16\epsilon_{\star}~, \\
n_s&=&1-6\epsilon_{\star}+2\eta_{\star}~,
\label{22}
\end{eqnarray}
All quantities with subscript `$\star$'  in the above equations are evaluated at a field value $\varphi=\varphi_{\star}$ that corresponds to a number of e-folds of the `visible' inflation, $N_{\star}\approx 30-65$: 
\begin{equation}
N_{\star}\simeq \frac{1}{M_P}\int_0^{\varphi_{\star}}\frac{d\varphi}{\sqrt{2\epsilon}}~.
\label{23}
\end{equation}  

In order to proceed with the actual calculations of the above observables, we plug 
Eqs. (\ref{12},\ref{13}) into Eq. (\ref{17}) and using Eq. (\ref{15}) we express the effective potential in terms of inflaton field $\varphi$ in the Einstein frame. Next, let us consider now the conformal limit where $\sigma_0\to 0$ and $\zeta_0 \to -1/12$. The latter limit implies that $\zeta$ evolves slowly, $\zeta \approx \zeta_0$. Assuming further, $\sigma_0^2 \sqrt{\frac{2\zeta_0}{1+12\zeta_0}}$ approaches to some constant $C$, the potential (\ref{17}) is well approximated by a potential which is linear in the inflaton field $\varphi$\footnote{A linear potential was obtained in a different limit of the non-minimally coupling in \cite{Kannike:2015kda}}:
\begin{equation}
V(\varphi)\approx \frac{162C}{\pi^2}M_P^3\varphi~.
\label{27}  
\end{equation}
The linear potential (\ref{27}) can be used to compute inflationary observables (\ref{18}-{23}). This immediately implies $\eta=0$ and hence the relation in Eq. (\ref{1}). In terms of observable efolds $N_{\star}$ the predictions read:
\begin{eqnarray}
\label{28}
n_s-1\approx -0.025\left(\frac{N_{\star}}{60}\right)^{-1}~, \\
r=0.0667\left(\frac{N_{\star}}{60}\right)^{-1}~.
\label{29}
\end{eqnarray}
$P_s\simeq 10^{-9}$ in turn implies $C\approx 5.5\cdot 10^{-12}\left(\frac{N_{\star}}{60}\right)^{-3/2} $. The predictions in Eq. (\ref{28}) are in a reasonable agreement with the most recent analysis of the cosmological data \cite{Huang:2015cke}, which suggests:
\begin{eqnarray}
\label{30}
n_s = 0.9669\pm 0040~~{\rm (68\% C.L.)}~,\\
\label{31}
r_{0.01} < 0.0685~~{\rm (95\% C.L.)}~,
\end{eqnarray}
for $\Lambda-$CDM$+r$ model. Further improvement of the accuracy of cosmological measurements will be critical for our scenario.

Note that for large $(\xi\to\infty)$ and small $(\xi\to 0)$ non-minimal couplings $n_s \gtrsim 1$, and thus the model is excluded by observation in these limits. 
 
\paragraph{Conclusion.}
We have proposed a new class of natural inflation models with hidden scale invariance realised via the dilaton field. A very generic Wilsonian potential with an arbitrary number of scalar fields contain a flat direction in the classical limit, which is lifted by quantum corrections. Thus inflation can naturally, without fine-tuning,  proceed when the inflaton field evolves along this direction. We find that in the conformal limit, the inflaton potential is linear, yielding to the specific predictions in Eqs. (\ref{28}) and (\ref{29}). While they are still in agreement with observations, more accurate cosmological measurements may turn critical in falsifying our scenario.

\paragraph{Acknowledgements.}
  
The work was supported in part by the Australian Research Council. AK was also supported in part by the Shota Rustaveli National Science Foundation (DI/12/6-200/13).


\begin{thebibliography}{99}

\bibitem{Guth:1980zm} 
  A.~H.~Guth,
  Phys.\ Rev.\ D {\bf 23}, 347 (1981).

\bibitem{Linde:1981mu} 
  A.~D.~Linde,
  Phys.\ Lett.\ B {\bf 108}, 389 (1982).

\bibitem{Albrecht:1982wi} 
  A.~Albrecht and P.~J.~Steinhardt,
  Phys.\ Rev.\ Lett.\  {\bf 48}, 1220 (1982).
  
\bibitem{Kazanas:1980tx} 
  D.~Kazanas,
  Astrophys.\ J.\  {\bf 241}, L59 (1980).
  
\bibitem{Starobinsky:1980te} 
  A.~A.~Starobinsky,
  Phys.\ Lett.\ B {\bf 91}, 99 (1980).
  
\bibitem{Sato:1981ds} 
  K.~Sato,
  Phys.\ Lett.\ B {\bf 99}, 66 (1981);
  Mon.\ Not.\ Roy.\ Astron.\ Soc.\  {\bf 195}, 467 (1981).

\bibitem{Mukhanov:1981xt} 
  V.~F.~Mukhanov and G.~V.~Chibisov,
  JETP Lett.\  {\bf 33}, 532 (1981)
  [Pisma Zh.\ Eksp.\ Teor.\ Fiz.\  {\bf 33}, 549 (1981)].

\bibitem{Ade:2015lrj} 
  P.~A.~R.~Ade {\it et al.} [Planck Collaboration],
  arXiv:1502.02114 [astro-ph.CO].

\bibitem{Huang:2015cke} 
  Q.~G.~Huang, K.~Wang and S.~Wang,
  arXiv:1512.07769 [astro-ph.CO].

\bibitem{Freese:1990rb} 
  K.~Freese, J.~A.~Frieman and A.~V.~Olinto,
  Phys.\ Rev.\ Lett.\  {\bf 65}, 3233 (1990).


\bibitem{Wetterich:1983bi} 
  C.~Wetterich,
  Phys.\ Lett.\ B {\bf 140}, 215 (1984); 

\bibitem{Bardeen:1995kv}
  W.~A.~Bardeen,
  FERMILAB-CONF-95-391-T.

\bibitem{Foot:2013hna} 
  R.~Foot, A.~Kobakhidze, K.~L.~McDonald and R.~R.~Volkas,
  Phys.\ Rev.\ D {\bf 89}, no. 11, 115018 (2014)
  [arXiv:1310.0223 [hep-ph]].

\bibitem{Kobakhidze:2014afa} 
  A.~Kobakhidze and K.~L.~McDonald,
  JHEP {\bf 1407}, 155 (2014)
  [arXiv:1404.5823 [hep-ph]].


\bibitem{GarciaBellido:2011de} 
  J.~Garcia-Bellido, J.~Rubio, M.~Shaposhnikov and D.~Zenhausern,
  Phys.\ Rev.\ D {\bf 84}, 123504 (2011)
  [arXiv:1107.2163 [hep-ph]].

\bibitem{Kallosh:2013hoa} 
  R.~Kallosh and A.~Linde,
  JCAP {\bf 1307}, 002 (2013)
  [arXiv:1306.5220 [hep-th]].

\bibitem{Salvio:2014soa} 
  A.~Salvio and A.~Strumia,
  JHEP {\bf 1406}, 080 (2014)
  [arXiv:1403.4226 [hep-ph]].
  
\bibitem{Ellis:2014gxa} 
  J.~Ellis, M.~A.~G.~Garcia, D.~V.~Nanopoulos and K.~A.~Olive,
  JCAP {\bf 1408}, 044 (2014)
  [arXiv:1405.0271 [hep-ph]].
  
\bibitem{Kannike:2014mia} 
  K.~Kannike, A.~Racioppi and M.~Raidal,
  JHEP {\bf 1406}, 154 (2014)
  [arXiv:1405.3987 [hep-ph]].

\bibitem{Csaki:2014bua} 
  C.~Csaki, N.~Kaloper, J.~Serra and J.~Terning,
  Phys.\ Rev.\ Lett.\  {\bf 113}, 161302 (2014)
  [arXiv:1406.5192 [hep-th]].
  
\bibitem{Kannike:2015apa} 
  K.~Kannike, G.~Hütsi, L.~Pizza, A.~Racioppi, M.~Raidal, A.~Salvio and A.~Strumia,
  JHEP {\bf 1505}, 065 (2015)
  [arXiv:1502.01334 [astro-ph.CO]].
  
  \bibitem{Ozkan:2015kma}
  M.~Ozkan and D.~Roest,
  arXiv:1507.03603 [hep-th].
  \bibitem{Kannike:2015fom}
    K.~Kannike, G.~Hütsi, L.~Pizza, A.~Racioppi, M.~Raidal, A.~Salvio and A.~Strumia,
    PoS EPS {\bf -HEP2015} (2015) 379.
\bibitem{Farzinnia:2015fka} 
  A.~Farzinnia and S.~Kouwn,
  arXiv:1512.05890 [hep-ph].
  
  \bibitem{Rinaldi:2015uvu}
  M.~Rinaldi and L.~Vanzo,
  arXiv:1512.07186 [gr-qc].

\bibitem{Okada:2010jf} 
  N.~Okada, M.~U.~Rehman and Q.~Shafi,
  Phys.\ Rev.\ D {\bf 82}, 043502 (2010)
  [arXiv:1005.5161 [hep-ph]].
  
\bibitem{Bezrukov:2007ep} 
  F.~L.~Bezrukov and M.~Shaposhnikov,
  Phys.\ Lett.\ B {\bf 659}, 703 (2008)
  [arXiv:0710.3755 [hep-th]].
  
\bibitem{Odintsov:1993rt} 
  S.~D.~Odintsov,
  Phys.\ Lett.\ B {\bf 306}, 233 (1993)
  [gr-qc/9302004].

\bibitem{Kannike:2015kda}
K.~Kannike, A.~Racioppi and M.~Raidal,
JHEP {\bf 1601} (2016) 035
doi:10.1007/JHEP01(2016)035
[arXiv:1509.05423 [hep-ph]].

\end{thebibliography}
\end{document}